\input{aipcheck}

\documentclass[
    ,final            
  ]
  {aipproc}

\layoutstyle{6x9}

\begin{document}

\title{Overview of progress in neutrino scattering measurements}


\author{M.~Sorel}{
  address={Instituto de Fisica Corpuscular (IFIC), CSIC and Universidad de
Valencia, Spain}
}

\begin{abstract}
 Recent progress in neutrino scattering experiments with few GeV
 neutrino beams is reviewed, focusing on new experimental input
 since the beginning of the NuInt workshop series in 2001. Progress
 in neutrino quasi-elastic scattering, resonance production,
 coherent pion production, scattering in the transition region between
 the resonance and deep inelastic regimes, and nuclear effects in
 neutrino-nucleus scattering, is discussed.
\end{abstract}

\maketitle


\section{Introduction}
Recent years have been marked by a wealth of new data on
neutrino-nucleus scattering with accelerator-based neutrino
beams. Input has been obtained mostly from experiments whose primary
goal is the study of neutrino oscillations, using near
detector data in long-baseline experiments and muon neutrino
interaction samples in experiments searching for small amounts of 
$\nu_{\mu}\to\nu_e$ and/or $\nu_{\mu}\to\nu_{\tau}$
transitions. In 
addition, valuable
data from test beam R\&D programs, and re-analyses of old
bubble chamber data, have recently become available.
Data in the 0.5$<E_{\nu}<$30 GeV neutrino energy range is discussed.
Experiments reporting progress in this range since 2001
include K2K near detectors, MiniBooNE,
NOMAD, the MINOS near detector, a 50-lt LAr TPC prototype exposed to 
the CERN WANF
neutrino beam (LAr50), as well as re-analyses of the BNL-7ft and 
Gargamelle (GGM) data. Key
parameters for these neutrino experiments are given in Tab.~\ref{tab:nuexps}.


Two cautionary remarks are in order. First, many of the results discussed
here are still preliminary, and may change in the near future. Second, direct
comparisons between experiments are sometimes difficult to perform; an
attempt has been made to explicitly give all assumptions made in
obtaining those comparisons.
\begin{table}
\begin{tabular}{ccccc}
\hline
  \tablehead{1}{c}{b}{Experiment}
  & \tablehead{1}{c}{b}{Flux-averaged\\energy (GeV)}
  & \tablehead{1}{c}{b}{Main nuclear\\target}
  & \tablehead{1}{c}{b}{Detector\\type}
  & \tablehead{1}{c}{b}{Neutrino\\interactions}  \\
\hline
MiniBooNE    & 0.8 & carbon & Cherenkov                       & $\sim 10^6$ \\
K2K-1KT      & 1.2 & oxygen     & Cherenkov                   & $\sim 10^5$ \\
K2K-SciFi    & 1.2 & oxygen     & segmented tracker           & $\sim 10^4$ \\
K2K-SciBar   & 1.2 & carbon & segmented tracker               & $\sim 10^4$ \\
BNL-7ft      & 1.6 & deuterium & bubble chamber               & $\sim 10^3$ \\
GGM          & 2.2 & propane & bubble chamber           & $\sim 10^3$ \\
MINOS-near   & 4.5 & iron   & magnetized tracking calorimeter & $\sim 10^6$ \\
NOMAD        & 24  & carbon    & spectrometer/calorimeter     & $\sim 10^6$ \\
LAr50        & 24  & argon     & TPC                          & $\sim 10^4$ \\
\hline
\end{tabular}
\caption{Characteristics of experiments reporting recent progress in
neutrino-nucleus scattering in the few GeV region.}
\label{tab:nuexps}
\end{table}

\section{\label{sec:qe}Quasi-Elastic Scattering}
Charged-current quasi-elastic (CCQE) scattering 
corresponds to the process $\nu_l n\to l^- p$, where 
$l=e,\mu ,\tau$. This is the dominant interaction channel
up to neutrino energies of about 1.5 GeV. In the Llewellyn Smith 
formalism \cite{Llewellyn Smith:1971zm}
employed to
describe this process, the differential cross-section $d\sigma /dQ^2$
depends mostly on two vector and one axial
vector form factors. The $Q^2$ dependence of the axial vector form factor
is typically assumed to have the dipole form,
$F_A(Q^2)\propto (1+Q^2/m_A^2)^{-2}$,
where $m_A$ is the CCQE axial mass,
consistent with current experimental accuracy.
Few percent deviations from the dipole form have
recently been observed in vector form factors using electron scattering data,
causing few 
percent differences in the CCQE cross section and axial mass extraction 
in recent analyses, compared to past ones where dipole vector
form factors were assumed. To first order, once
electron scattering and nuclear $\beta$ decay experimental input
is used, the task of measuring
the neutrino-nucleon CCQE cross section can be recast as the
task of measuring the only free parameter left in the 
model: the CCQE axial mass. 

Neutrino interactions in K2K-SciFi have been used 
to extract $m_A$ \cite{Gran:2006jn}, by fitting the shape 
of the
$Q^2$ distribution in the 1 track and 2 track QE-enriched charged-current
(CC) samples. In the fit, the 2 track nonQE-enriched CC sample is
also included, to constrain the background normalization. The $Q^2$ fit is
performed in separate neutrino energy bins, to constrain neutrino flux
predictions. Also, the fit is performed for $Q^2>0.2$ GeV$^2$
only, to avoid large uncertainties due to nuclear effects. Overall,
the total sample used corresponds to about 7,000
interactions. The result obtained is: $m_A=(1.20\pm 0.12)$ GeV.
A similar analysis has been performed 
using neutrino data from K2K-SciBar.
A preliminary axial mass measurement was presented at this workshop
\cite{k2kscibarma}, corresponding to $m_A=(1.144\pm 0.077\hbox{(fit)}
^{+0.078} _{-0.072}\hbox{(syst)})$ GeV.

MiniBooNE has also measured the CCQE axial mass parameter
\cite{miniboonema}, by requiring contained
CC events with a single decay electron tag, correlated in
space with the muon track end-point. Overall, about 200,000 events
with a 74\% estimated CCQE purity are used. A
fit to the $Q^2$ shape is performed, to measure both
the axial mass and a parameter controlling the strength of the Pauli
suppression in a relativistic Fermi gas model for the carbon nucleus
\cite{Smith:1972xh}.
A good data/Monte Carlo agreement in the CCQE kinematic distributions is
achieved after tuning these two parameters in the simulation. The
MiniBooNE axial mass result is: $m_A=(1.23\pm 0.20)$ GeV.


The axial mass results from K2K-SciFi, K2K-SciBar, and
MiniBooNE agree with each other,
but appear to be higher than the
historical world average \cite{Bernard:2001rs}. Previous
quasi-elastic $m_A$ results had been
obtained from neutrino and antineutrino scattering with
deuterium/hydrogen \cite{ma_deuteriumhydrogen}, 
propane/freon \cite{ma_propanefreon},
iron \cite{ma_iron}, and carbon \cite{ma_carbon} targets.

Charged-current quasi-elastic interactions
have been studied in NOMAD \cite{Lyubushkin:2006dv} by
requiring two tracks, one 
identified as $\mu^-$ and the other consistent with a proton,
an invariant hadronic mass $W<1.76$ GeV, and a CCQE-like signature
based on a 3-dimensional event likelihood. Overall, about 8,000
events are selected, with an estimated CCQE purity of 71\%.
The absolute CCQE cross section as a function of energy is 
normalized to deep inelastic events. The preliminary NOMAD
result is $\sigma (\nu_{\mu}n\to\mu^-p)=
(0.72\pm 0.01)10^{-38}$ cm$^2$ for $3<E_{\nu}<100$ GeV,
where the error quoted is statistical-only. Unlike
the results described above, this
measurement is about 20\% smaller than the world average value,
suggesting a correspondingly smaller axial mass value.
The ongoing systematic uncertainty evaluation,
expected to be dominated by nuclear effects,
is needed before any definitive statement can be made.

Closely related to CCQE scattering, a preliminary differential 
cross section for neutral-current (NC) elastic scattering, 
$\nu_{\mu}N\to\nu_{\mu}N$,
has been presented at this workshop by MiniBooNE
\cite{cox}. Results are consistent with past findings
\cite{ma_carbon}.

\section{\label{sec:res}Resonance Production}
Pions can be produced in neutrino interactions via the excitation, and
subsequent decay, of resonances. The NC process is:
$\nu_{\mu}N\to\nu_{\mu}N^{\ast}\to\nu_{\mu}N'\pi$,
where $N$ and $N'$ are nucleons, and $N^{\ast}$ is a hadronic resonance. 
A similar process holds
for resonant CC interactions. Typically, resonances with
hadronic masses in the range $1.08<W<1.4-2.0$ GeV are considered.
This interaction channel is the dominant one in the
$1.5< E_{\nu}< 2.5$ GeV neutrino energy range, approximately.
Resonance production is generally modeled according to the Rein and
Sehgal formalism \cite{Rein:1980wg}. 
The transition form factor appearing in the
resonance production amplitude is assumed to have a dipole form, with
a $Q^2$ dependence controlled by a single pion axial mass parameter. 

Since NuInt01, re-analyses of BNL-7ft \cite{Furuno:2003ng} and GGM
\cite{ggmncpi0} resonance production data have been performed. Good
agreement with Rein-Sehgal expectations was found in
both cases, for single pion axial masses in the 1.0-1.2 GeV range.
The GGM re-analysis provided the first absolute cross section
measurement of resonant NC $\pi^0$ production in the
few GeV neutrino energy range, $\nu_{\mu}N\to\nu_{\mu}N\pi^0$.
 
Neutral-current $\pi^0$ production interactions have been
observed with the K2K-1KT detector \cite{Nakayama:2004dp}, by 
requiring two electron-like
Cherenkov rings with an invariant mass in the $85<m_{\gamma\gamma}
<215$ MeV range. About 2,500 events were identified,
with an estimated purity of 71\%.
The cross section obtained,
normalized to the CC inclusive one, is:
$\sigma (\hbox{NC}\pi^0)/\sigma (\hbox{CC}) = (0.064\pm 0.001
\pm 0.007)$, for a neutrino energy $E_{\nu}\sim 1.5$ GeV.
The K2K-1KT $\pi^0$ momentum distribution
shows reasonable agreement with predictions.

Resonant NC $\pi^0$ production has also been studied by MiniBooNE.
A preliminary absolute cross section of
$\sigma (\nu_{\mu}N\to\nu_{\mu}N\pi^0)=
(1.28\pm 0.11\pm 0.43)10^{-38}$ cm$^2$/CH$_2$
was reported in \cite{Raaf:2005up} for a
neutrino energy of $E_{\nu}\sim 1.3$ GeV. This measurement
assumes a neutrino flux extracted via MiniBooNE CCQE data,
where $m_A=1.03$ GeV was used. Several
analysis updates occurred since then \cite{link}. The updated
selection requires no decay electrons, $e/\mu$ and $e/\pi^0$ likelihood
ratios favoring the electron and $\pi^0$ hypotheses, respectively,
and two Cherenkov rings with an invariant mass in the
$80<m_{\gamma\gamma}<200$ MeV range. Approximately 20,000
NC $\pi^0$ candidates are selected, with high
($>90$\%) purity, allowing to measure the rate
of $\pi^0$ production. The observed $\pi^0$ momentum spectrum is
$\sim$20-30\% softer than predictions \cite{Casper:2002sd}.


The resonant NC $\pi^0$ production cross sections by GGM
and MiniBooNE, and the cross section ratio result by K2K-1KT multiplied
by the K2K-1KT prediction for the CC inclusive cross
section, can be compared with each other, and with past data obtained 
with the ANL-12ft bubble chamber \cite{anlncpi0}. The results are
found to be reasonably consistent with each other. 

In addition to $\nu_{\mu}N\to\nu_{\mu}N\pi^0$ interactions, 
preliminary results on the $\nu_{\mu}N\to\mu^-N\pi^+$ channel
have also been obtained by MiniBooNE
\cite{Wascko:2006tx}. About 44,000 event candidates
are selected by requiring two decay electrons, corresponding to about a
85\% purity. The neutrino energy can be reconstructed 
from the observed muon kinematics, and the
$\nu_{\mu}N\to\mu^-N\pi^+$ rate is normalized to the
$\nu_{\mu}n\to\mu^-p$ one. By multiplying
this measured cross section ratio by the CCQE cross section 
prediction
($m_A=1.03$ GeV), a $\nu_{\mu}N\to\mu^-N\pi^+$ cross section as
a function of neutrino energy can be inferred. The central value
result is $\sim 25$\% lower than predictions, but uncertainties
are estimated to be of comparable size as the difference between 
data and predictions. This result is in the process of being
updated.

Additional results on CC resonance production with the
K2K-SciBar detector have been presented for the first time during
this workshop, concerning both the $\nu_{\mu}N\to\mu^-N\pi^+$
and the $\nu_{\mu}n\to\mu^-p\pi^0$
channels \cite{k2kscibar_resonant}.

\section{\label{sec:coh}Coherent Pion Production}
In addition to resonant pion production, neutrinos can produce pions
also by interacting coherently with the nucleons bound in a nucleus.
The cross section for this process is expected to be smaller than
resonant pion production, up to $\sim$20\% for neutrino energies of
about 1 GeV, but with a distinct signature, consisting of a 
forward-scattered pion and no nuclear break-up. Both CC
and NC modes are possible, $\nu_{\mu}A\to\mu^-A\pi^+$ and
$\nu_{\mu}A\to\nu_{\mu}A\pi^0$, respectively. Neutrino and antineutrino 
coherent cross sections are expected to be similar \cite{alvarez}. 

The first result on CC coherent pion production
was obtained with the K2K-SciBar detector \cite{Hasegawa:2005td}.
Charged-current coherent pion candidates are selected by requiring 
a CC interaction with two tracks,
one muon and one $\pi^+$-like track, low vertex activity, and low
momentum transfer ($Q^2<0.1$ GeV$^2$). Control samples are used to
tune the momentum scale, the nQE/QE cross section ratio, and the
strength of nuclear effects. Overall, 113 events were selected,
consistent with background-only events. Based on this result, 
an upper limit
on the CC coherent pion cross section, normalized to
the CC inclusive one, of $\sigma (\nu_{\mu}A\to\mu^-A\pi^+)/
\sigma (\nu_{\mu}N\to\mu^-X)<0.60\cdot 10^{-2}$ is obtained, 
at 90\% confidence
level and for a mean neutrino energy of $E_{\nu}\simeq 1.3$ GeV.

Neutral-current coherent pion production has been studied
by MiniBooNE \cite{link} using a sample of
about 30,000 events, selected by requiring no decay electrons,
$e/\mu$ and $e/\pi^0$ likelihood ratios favoring the electron and
$\pi^0$ hypotheses, respectively, and $m_{\gamma\gamma}>50$ MeV. 
A two-dimensional fit in the $(m_{\gamma\gamma},
E_{\pi^0}(1-\cos\theta_{\pi^0}))$ observables was performed to extract
the coherent, resonant, and background fractions in the sample. 
A clear evidence for a non-zero coherent fraction in the
$\pi^0$ sample
was found, corresponding to: 
$N_{\hbox{coh}}/(N_{\hbox{coh}}+N_{\hbox{res}})=
(18.0\pm 1.2\pm 1.0)$\% for a neutrino energy of $E_{\nu}\simeq 1.1$
GeV. MiniBooNE also presented at this workshop evidence for antineutrino 
coherent $\pi^0$ production in the forward scattering region
\cite{nguyen}.


The MiniBooNE neutrino NC coherent pion production measurement, 
converted into an absolute cross section using \cite{Raaf:2005up},
is consistent with past results on $\nu_{\mu}A\to\nu_{\mu}A\pi^0$
and $\bar{ \nu}_{\mu}A\to\bar{\nu}_{\mu}A\pi^0$
obtained by the
Aachen-Padova \cite{Faissner:1983ng} and GGM \cite{Isiksal:1984vh}
experiments at $E_{\nu}\simeq 2$ GeV and 3.5 GeV,
respectively. In this comparison,
an $A^{2/3}$ dependency of the coherent pion cross section was assumed
\cite{alvarez}. A comparison between CC and NC 
coherent pion production cross section results can also be performed,
by assuming $\sigma (\nu_{\mu}A\to\mu^-A\pi^+)\simeq
2\sigma (\nu_{\mu}A\to\nu_{\mu}A\pi^0)$
\cite{alvarez}. This comparison indicates some tension between the
positive observation of NC coherent pion production
by the MiniBooNE, Aachen-Padova and GGM experiments on one side,
and the upper limit on CC coherent pion production reported
by K2K-SciBar on the other side. In the comparison, the 
K2K-SciBar result on $\sigma (\nu_{\mu}A\to\mu^-A\pi^+)/
\sigma (\nu_{\mu}N\to\mu^-X)$ is multiplied by the
CC inclusive cross section value predicted for K2K.
A variety of models on coherent pion production have been proposed
\cite{coherenttheory},
often yielding very different predictions; more experimental
input appears to be necessary to guide the theory.

\section{\label{sec:dis}From Resonance Production to Deep Inelastic Scattering}

As the neutrino energy increases, neutrino-nucleus interactions
can resolve the composite structure of the nucleon targets, and
more hadronic final states become
kinematically allowed. Multiple pion
production becomes an important process for neutrino
energies above 2 GeV. For $W>2$ GeV hadronic masses, it is customary 
to use the deep inelastic
scattering (DIS) formalism to describe neutrino-nucleus interactions
\cite{Conrad:1997ne}.
This is the dominant type of interaction above
about 3 GeV neutrino energy.
A smooth transition from the resonance production to the DIS regimes
is typically modeled via duality-inspired models \cite{Bodek:2003wd}.
Typical observables in CC DIS are the
muon energy $E_{\mu}$, muon direction $\theta_{\mu}$, and the total 
hadronic energy $E_h$. Differential
cross sections are typically expressed in terms of the Bjorken
scaling variable $x\equiv Q^2/(2m_NE_h)$ and inelasticity
$y\equiv (E_h-m_N)/E_{\nu}$, where $Q^2=4E_{\nu}E_{\mu}\sin^2(\theta_{\mu}/2)$
is the momentum transfer, $E_{\nu}$ is the neutrino energy, $m_N$ is
the nucleon mass. 
The DIS cross section is 
typically expressed in terms of
the nucleon structure functions $F_2(x,Q^2)$, $xF_3(x,Q^2)$, and $R_L(x,Q^2)$.

The MINOS near detector already collected large samples of DIS
($W>2$ GeV) and transition region ($1.4<W<2$ GeV) interactions. Preliminary
event distributions as a function of $x$ and $y$ for the DIS sample
show reasonable agreement with predictions \cite{naples}. One of the 
MINOS near-term goals 
is to extract the $F_2$ and $xF_3$ structure
functions for neutrino-iron scattering, using event distributions
of this type
and a neutrino flux extraction obtained for $E_{\nu}>5$ GeV, $E_h<1$ GeV,  

NOMAD recently measured the CC differential
cross section $d^2\sigma/(dxdy)$ in neutrino-carbon interactions in
the $6<E_{\nu}<300$ GeV neutrino energy range as a function
of $E_{\nu}$ \cite{Petti:2006tu}.
The selection requires muon identification, a muon (hadronic) energy
greater than 2.5 (3) GeV, and $Q^2>1$ GeV$^2$. The cross section 
normalization is obtained using
the world average value for
$40<E_{\nu}<200$ GeV. This is the first
measurement of the inelastic CC neutrino cross section on
carbon at large ($Q^2\sim 13$ GeV$^2$) momentum transfers. 

Overall, the MINOS and NOMAD experiments cover regions of
phase space at high $x$ and low/medium $Q^2$, allowing to perform
structure functions measurements that are complementary to those
performed by charged lepton scattering experiments. 

\section{\label{sec:nuclear}Nuclear Effects}
Most interactions in current- and next-generation
neutrino experiments occur with target nucleons bound in nuclei.
Nuclear effects affecting neutrino-nucleus interactions can
be divided into three categories \cite{Llewellyn Smith:1971zm}:
first, Fermi motion and binding
energy of target nucleons, changing the interaction kinematics;
second, Pauli suppression of the phase space
available to final state nucleons, causing a $Q^2$-dependent
cross section reduction; third, final state interactions (FSI)
inside the nucleus,
changing the composition and kinematics of the hadronic
part of the final state. A relativistic Fermi gas model for the target 
neutrons and protons \cite{Smith:1972xh} is generally used to model
Fermi motion and Pauli suppression.
A variety of models tuned on pion and proton scattering data
are available to simulate FSI effects.
Depending on experimental energy thresholds and target nuclei
involved, nuclear
de-excitation via gamma ray emission may also be relevant.

Nuclear de-excitation has been observed
with the K2K-1KT detector \cite{Kameda:2006zn}.
About 40\% of the neutrino
interactions with nucleons bound in oxygen are
accompanied by $\sim$6 MeV gamma ray emission from nuclear
de-excitation. This signature potentially allows
the study of NC elastic scattering
in water Cherenkov detectors. About 3,000 gamma candidate events 
with an estimated
58\% $\nu_{\mu}N\to\nu_{\mu}N$ purity were selected, by
requiring low PMT hit multiplicity, event containment, and
a single Cherenkov ring hit topology. A $\sim$6 MeV visible energy
peak has been unambiguously observed. The
observed-to-predicted event rate ratio in the sample was
measured to be: $1.23\pm 0.04\pm 0.06$, where the systematic
error includes detector uncertainties only. In the
comparison, the predicted event rate is normalized
to the K2K-1KT CC inclusive measurement.

Low-$Q^2$ neutrino interactions, the most affected by
nuclear effects, have been recently studied by
the MiniBooNE and K2K experiments. Early analyses of various low-$Q^2$ samples
at both MiniBooNE and K2K showed a deficit with respect to
predictions for $Q^2<0.2$ GeV$^2$. Since then, the two experiments
have followed different approaches to tune low-$Q^2$ predictions.
The MiniBooNE Collaboration introduced an extra degree of freedom
in the relativistic Fermi gas model to control the strength of
the Pauli suppression, explaining the deficit via
nuclear physics arguments \cite{miniboonema}.
The K2K Collaboration found that
most (if not all) of the discrepancy is eliminated by assuming
no CC coherent pion production, explaining the deficit
via neutrino interaction arguments \cite{Hasegawa:2005td}.

Nuclear effects have also been studied in a 50-lt LAr TPC prototype
exposed to the CERN multi-GeV wide-band beam \cite{Arneodo:2006ug}.
The detector prototype
used NOMAD as a muon spectrometer. A ``golden'' CCQE sample of 86
events was selected, with an estimated purity of about
80\%. Among the observables that were studied,
the missing transverse momentum is particularly sensitive to nuclear
effects such as Fermi motion, proton re-scattering and pion absorption
inside the argon nucleus. Evidence for nuclear effects
beyond Fermi motion and Pauli suppression was clearly observed.

In general, a great effort has been devoted in recent neutrino
scattering analyses not only to correct for nuclear effects, but
also to try to quantitatively evaluate the uncertainties associated
with nuclear corrections. This represents a
recent, encouraging trend. 
Experiments typically tend to be conservative in quoting nuclear
uncertainties, also to ``cover'' possible nuclear model
deficiencies, such as the ones associated with the simple
relativistic Fermi gas model that is commonly used.  

\section{Conclusions}

Recent progress in neutrino scattering measurement has been
reviewed, focusing on new experimental input since the
beginning of the NuInt workshop series in 2001. Several new
cross section results, spanning all relevant 
interaction channels, have become available, as well as
new studies of nuclear effects in neutrino-nucleus interactions.
Recent event samples tend to have much higher statistics
than previous data sets, allowing a more comprehensive
study of differential cross sections. Despite
the large statistics, recent neutrino cross section
results are often still 
characterized by large uncertainties. Even though this
may appear deceiving, large uncertainties tend to represent more
accurately than in the past our understanding of systematic effects. 
Nevertheless, 
examples have been given of recent results that are not
consistent with each other and with past ones,
pointing to either non-understood experimental biases,
or deficiencies in the models used to analyze and interpret the data.
These inconsistencies need to be resolved for an
accurate understanding of neutrino-nucleus 
interactions in the few GeV region.

Fortunately, the future of the field is bright. First,
a wealth of new neutrino experiments is expected
to provide data in the near future, such as NuSNS
at Oak Ridge,
SciBooNE and MINER$\nu$A at Fermilab, as well as the
near detectors for the MINOS and NO$\nu$A experiments
at Fermilab, and for the T2K experiment at J-PARC \cite{futureexps}.
Second, the neutrino scattering community has established 
strong synergies with the nuclear physics and charged lepton
scattering communities, and a close collaboration
between theoretical and experimental physicists, as widely
demonstrated by this workshop series.


\begin{theacknowledgments}
The author would like to thank the organizing committee and
participants for the stimulating atmosphere at 
the NuInt07 workshop, and J.~Alcaraz, Z.~Pavlovic and G.~Zeller
for useful suggestions and information. 
The author is supported by a 
Marie Curie Intra-European Fellowship within the 6th European 
Community Framework Program.
\end{theacknowledgments}

\end{document}